\documentstyle[11pt,a4]{article}
\begin{document}
\begin{center}{\large
 Gravitational Correction in Neutrino Oscillations
}\end{center}
\begin{center}{ Yasufumi Kojima \footnote{
E-mail address: kojima@theo.phys.sci.hiroshima-u.ac.jp}  \\
       Department of Physics, Hiroshima University,  \\
         Higashi-Hiroshima 739, JAPAN              
 }\end{center}
\begin{abstract}
We investigate the quantum mechanical oscillations of neutrinos propagating 
in weak gravitational field.
The correction to the result in the flat space-time is derived.
\end{abstract}

   Recently, the neutrino oscillation phase due to the presence 
of gravity is much discussed 
\cite{Ahluwalia},  \cite{BHM}, \cite{Ahluwalia2}.
If the neutrinos are massive particles and mixed, the neutrino
oscillation between different flavor states occurs.
Suppose that neutrinos are created as weak flavor eigenstate, say,
at ${\vec r}_A$, and propagate to ${\vec r}_B$.
The state is a linear superposition of mass eigenstates and each phase of 
mass eigenstate evolves in different way.
As a result, the mixing phase angle for the relativistic neutrinos 
propagating in  a flat space is given by
$\varphi _0 = \Delta m^2 L /( 4 \hbar E ) $,  
where $E $ is energy, $ L = |{\vec r}_B -{\vec r}_A| $,
and $ \Delta m^2 = m_1^2 - m_2^2$. 
(See e.g., \cite{Moh}.)

Gravitational effect was not seriously studied so far.
Gravity can be eliminated by choosing appropriate inertial frame locally.
The propagating distance of the neutrinos is so long in some case,
that gravitational effect may become important.
Ahluwalia and Burgard \cite{Ahluwalia}
considered the gravitational effect on the neutrino oscillation.
They showed that
the external weak gravitational field of a star with mass $M$ adds a new
contribution to the phase difference, denoted by
$\varphi _G = - \Delta m^2 L  \langle \phi  \rangle /( 4 \hbar E ) $,  
where $ \langle \phi  \rangle  $ 
is defined by the average of gravitational potential over 
the semi-classical path, i.e.,
$ \langle \phi  \rangle   = -( \int _{\vec r_A}  ^{\vec r_B} 
dL GM/r ) /L$.
This gravitationally induced phase can be estimated as
$\varphi _G = - \langle \phi  \rangle \varphi _0 $.
The phase becomes to the extent of roughly $20 \%$
of $\varphi _0 $, near neutron stars.
They suggested that the new oscillation phase may be significant 
effect on the supernova explosions,
since the extremely large fluxes of neutrinos are produced with 
different energies corresponding to the flavor states.

The gravitationally induced oscillation phase may have the important 
astrophysical consequences.
However, their derivation and even the definitions such as energy were
not clear in their original paper.
Bhattacharya, Habib and Mottola \cite{BHM}
critically re-examined the quantum mechanical phase mixing. They 
calculated the phase difference for radially propagating particles, and 
found that the term of $ \Delta m^2  L \langle \phi \rangle /(\hbar E) $
$ \sim GM \Delta m^2  \log(r_B/r_A)/(\hbar E) $
is canceled out.  They showed that the possible gravitational 
effect appears at the higher order, $ \Delta m^4 / E^3 $,
and that the phase difference for radially
propagating particles is 
$ G M \Delta m^4 $ $ \log(r_B/r_A)$ $/(4 \hbar E^3) $.
Numerically its magnitude is equal to 
$ \sim 10 ^{-9} $ $(M/M_\odot)$ 
$(\Delta m^4 / {\rm eV}^4)$ $(E/ {\rm MeV})^{-3} $ $\log(r_B/r_A) $,
which is completely negligible in typical
astrophysical applications.
Therefore, the conclusion of Ahluwalia and Burgard \cite{Ahluwalia}
seems to be incorrect for radially propagating case.
Natural question is what happens in more general case.
Does the term of $ G M \Delta m^2 / E $ always disappear?

Furthermore, several authors have discussed the possibility
of the violation of equivalence principle in the neutrino oscillation,
e.g.,\cite{IMY}, \cite{HLP} and reference therein.
If the universality of the gravitational couplings to
different flavors breaks down, additional phase difference appears.
It is, therefore, an important matter to understand the modification
of the neutrino oscillation phase due to the presence of gravity, 
even within the  equivalence principle.
This question must be settled first.
In this paper, we will consider the quantum mechanical
phase mixing of the neutrinos propagating 
in the weak gravitational field in detail.

Let us consider a weak flavor eigenstate
$ | \nu _\alpha \rangle $  $(\alpha = e, \mu, $ or $\tau)$,
created at $ { \vec r}_A$  with a certain energy.
The value can be specified  by the energy at infinity $E$.
The  flavor eigenstate is a linear superposition of mass eigenstates,
which are represented by   $ | \nu _a \rangle$,
\begin{equation}
  | \nu _\alpha \rangle  = \sum _a U _{ \alpha a }
  | \nu _a \rangle  ,
\end{equation}
where  $ U _{ \alpha a } $  is unitary mixing matrix.
The different mass eigenstates propagate with different velocity.
The classical paths deviate reciprocally in the gravitational field.
We use wave packet formalism (\cite{Kay}, \cite{GKL})
to examine the evolution of mass eigenstates.
We assume that each mass eigenstate is created with the same energy
as the Gaussian wave form in momentum space.
The direction of the momentum is the same, but
the mean value is different due to different mass.
The wave packet for mass eigenstate, $a$ can be written 
in coordinate space as
\begin{equation}
  \Psi _a ( {\vec x}, t) =
( 2 \pi \sigma^2) ^{ -3/4} \exp \left ( { i S_a \over \hbar} \right )
       \exp \left ( - { | {\vec x} - {\vec z}_a |^2 \over 4 \sigma ^2} 
       \right ) ,
\end{equation}
where $ S$ is phase function and should satisfy 
$ g^{\mu \nu } S, _\mu S_\nu + m^2 = 0 $
in geometrical optics limit.
The size of the wave packet, $ \sigma $ is assumed to be much smaller 
than the curvature radius of the external gravitational field.
The center of the wave packet, $ {\vec z}_a(t)$ 
is determined by classical geodesic.

The quantum mechanical transition probability from
$ | \nu _\alpha \rangle$  to $ | \nu _\beta \rangle$
is
\begin{eqnarray}
  P_{\alpha  \to \beta } ( {\vec x}, t) &=& 
   | \sum _a U^* _{ \beta a }
          \Psi _a ( {\vec x}, t) U _{ \alpha a } |^2
\nonumber \\
       &=&
( 2 \pi \sigma^2) ^{ -3/2}
\sum _{a b} U _{ \beta a } U^*  _{ \alpha a }
            U _{ \beta b } U^* _{ \alpha b }
 \exp \left ( { i \over \hbar} ( S_a - S_b) \right ) F ,
\end{eqnarray}
where
\begin{eqnarray}
  F( {\vec x}, t) &=&
   \exp \left ( - { 1 \over 4 \sigma ^2} ( 
| {\vec x} - {\vec z}_a |^2  + | {\vec x} - {\vec z}_b |^2 ) 
 \right )
\nonumber \\
       &=&
      \exp \left ( - { 1 \over 2 \sigma ^2}  
| {\vec x} - {1 \over 2} ( {\vec z}_a + {\vec z}_b) |^2   \right )
 \exp \left ( -  { 1 \over 8 \sigma ^2}  
 | {\vec z}_a  - {\vec z}_b |^2 
 \right ).
\end{eqnarray}
The function $ F$ has a peak at
$ ( {\vec z}_a + {\vec z}_b )/2 $,
so that the phase factor can be evaluated by taking the dominant 
contribution in the stationary point.
The factor  
$ \exp  ( - | {\vec z}_a  - {\vec z}_b |^2 /( 8 \sigma^2) )$
represents that the coherence will be lost if two
orbits separate much more than the size of the wave packet.
The unitary matrix for the mixing between two generations is
parameterized by $ \theta $.
If the deviation of the orbit is negligible, then we have
\begin{equation}
  P_{\alpha  \to \beta } ( t) = \sin ^2 (2 \theta) \sin ^2 ( \varphi), 
\end{equation}
where $ \varphi $ is the phase angle between two states,
$ \varphi = - \Delta S /( 2 \hbar) $
$ = (S_2 -S_1) /(2 \hbar)$.
In the opposite case, we have time-averaged one,
\begin{equation}
  P_{\alpha  \to \beta } = { 1 \over 2 } \sin ^2 (2 \theta) . 
\end{equation}

We now estimate the split of the orbit, $  {\vec z}_1 - {\vec z}_2 $,
and the phase difference  $ \Delta S = - 2 \hbar \varphi $,
between two mass eigenstates.
The classical trajectory of a particle with mass $m$ 
around non-rotating star with mass $M$ can be described 
by the energy at infinity $E$ and angular momentum $L$.
The orbital plane may be chosen as the equatorial plane of the 
spherical coordinate.
Then the trajectory leaving at $t=0$ from $(r, \chi) = (r_A, 0 ) $
can be written as
\begin{equation}
  \chi = \int _{r_A} ^r 
    {1 \over  \sqrt{B} } { L \over r^2} dr ,
\end{equation}
\begin{equation}
  t = \int _{r_A} ^r 
    {E \over  \sqrt{B} } 
\left ( 1 - { 2GM \over r}  \right ) ^{-1} dr ,
\end{equation}
where
\begin{equation}
  B = E^2 - \left ( 1 - { 2GM \over r}  \right )
          \left ( m^2 + { L^2 \over r^2 }  \right ) .
\end{equation}
We will consider unbounded orbit, i.e., $ E \gg m$.
It is convenient to use the turning point, $ r_0$,
of the radial direction to calculate the above integrals.
Eliminating $L$ and expanding by $G$, we have
\begin{equation}
  \chi = \left[ - \sin ^{-1} { r_0 \over r }
    + {GM  \over  1-q } {2r +r_0 \over  r r_0 } 
    \sqrt{ r-r_0 \over r+r_0 }
    - {qGM  \over  1-q } {\sqrt{ r^2 -r_0^2}  \over  r r_0 } 
  \right] _{r_A} ^r ,
\label{eqchi}
\end{equation}
\begin{equation}
  \sqrt{1 -q} t = \left[  \sqrt{ r^2 -r_0^2 }
    + {GM  \over  1-q } \sqrt{ r-r_0 \over r+r_0 }
    + {(2-3q)GM  \over  1-q } \log \left(
 { r + \sqrt{ r^2 -r_0^2 } \over  r_0 } \right) 
  \right] _{ r_A} ^r ,
\label{eqtm}
\end{equation}
where  $ q = m^2 /E^2$, and
$[ f(r)] ^r _{r_A} = f(r) -f(r_A)$.
The first terms in eqs. (\ref{eqchi}) and (\ref{eqtm})
represent straight lines in the spherical coordinate.
The first order terms in $G$ are well known effects on
the propagation of massless particles.
The second term in (\ref{eqchi}) is relevant to deflection of light, 
and the logarithmic term in  (\ref{eqtm})  is relevant to the time delay 
in the radar echo experiment.
(See, e.g., \cite{Wein}.)

The phase $S$ is given by the four momentum, $ p_\mu$
conjugate to $x^\mu$ as,
\begin{equation}
  S = \int p_\mu dx^\mu = \int ( -E dt + p_i dx^i ).
\end{equation}
Both mass eigenstate are assumed to have the same conserved energy $E$, 
so that the first term $ \int E dt $ is canceled out.
\footnote[1]{
If the mixed state is a superposition of states with different energy,
there is additional phase difference,
$ \Delta S = - \Delta E t = - \Delta m^2 t /(2 E).$
We have the logarithmic term in $ \Delta S$ by  eq.(\ref{eqtm}),
in converting from the coordinate time to the propagating distance.
We will not consider such a term from the transformation
between $t$ and $r$ hereafter.}
We only calculate the contribution from the second term.
The phase function can be evaluated for the classical path of 
the particle with mass $m$ and energy $E$ as
\begin{eqnarray}
  S  &=&  \sqrt{ E^2 -m^2 } \times
\nonumber \\
     &&
\left[ \sqrt{r^2 -r_0 ^2}  + GM \left(
  { 2-q \over 1-q} \log \left( { r+ \sqrt{ r^2 -r_0 ^2} \over r_0 }
\right) + {1 \over 1-q} \sqrt{ { r-r_0 \over r + r_0}}
-2 { \sqrt{ r^2 - r_0 ^2} \over r } \right)
\right] _{ r_A} ^r .
\end{eqnarray}
Expanding by $ m^2/E^2 $ for the relativistic particles, we have
\begin{eqnarray}
 S  & \simeq &  E \left[ 
\sqrt{r^2 -r_0 ^2}   +GM 
\left(2 \log \left({ r + \sqrt{ r^2 - r_0 ^2} \over r_0 } \right)
+\sqrt{ { r-r_0 \over r + r_0}}
-2 { \sqrt{  r^2-r_0^2 } \over r } \right)
\right] _{ r_A} ^r 
\nonumber \\
     && - { m^2 \over 2 E } \left[ 
\sqrt{r^2 -r_0 ^2}   -GM \left(
\sqrt{ { r-r_0 \over r + r_0}}
+2 { \sqrt{  r^2-r_0^2 } \over r } \right) 
\right] _{ r_A} ^r 
\nonumber \\
     && - { m^4 \over 8 E^3 } \left[ 
\sqrt{r^2 -r_0 ^2}   -GM \left(
2 \log \left({ r + \sqrt{ r^2 - r_0 ^2} \over r_0 } \right)
+3\sqrt{ { r-r_0 \over r + r_0}}
+2{ \sqrt{  r^2-r_0^2 } \over r } \right) 
\right] _{ r_A} ^r 
\nonumber \\
&& + \cdots .
\label{exs}
\end{eqnarray}
It is clear that the logarithmic term never appears
in order $G M m^2/ E $, but  in order  $G M m^4/ E^3 $.
However, there is still another term in order $G M m^2/ E $.
We only consider the gravitational effect up to 
$G M m^2/ E $,  and neglect the higher order terms hereafter.
The particle with $ m^2 \pm \Delta m^2 /2 $
arrives at $( r \pm \Delta r, \chi \pm \Delta \chi /2) $
at the same coordinate time $t$.
The split of the orbit can be calculated from
eqs. (\ref{eqchi}) and (\ref{eqtm}).
The phase difference $ \Delta S $ can be calculated from eq.(\ref{exs})
at $r$. We explicitly show the differences in the path and phase due to
mass square difference for the relativistic particles emitted
to the radial and azimuthal directions, i.e.,
${\dot \chi } = 0 $ and ${\dot r } = 0 $ initially.
The turning points for these orbits can be chosen as $r_0 = 0$,
and $r_0 = r_A $, respectively.

\vspace{0.5cm}
\noindent
(1){\it radial orbits}

The orbital deviation and phase difference 
in the radial direction are given by,
\begin{equation}
  \Delta r \simeq - { \Delta m^2 \over 2E^2 } \left\{
1 - {2GM  \over  r}  \right\} L,
\end{equation}
\begin{equation}
  \Delta S \simeq - { \Delta m^2 \over 2E } L,
\end{equation}
where $L$ is the path length in flat space, $ L = r - r_A$.

\vspace{0.5cm}
\noindent
(2){\it transverse  orbits}

For the particles emitted transversely, we have
\begin{equation}
  \Delta r \simeq - { \Delta m^2 \over 2E^2 } \left\{
1 + {4GM  \over  r+ r_A}  - {3GM  \over  r}  
\right \} { L \over r } L,
\end{equation}
\begin{equation}
  r \Delta \chi  \simeq - { \Delta m^2 \over 2E^2 } \left\{
1 + {3GM  \over  r_A}  + {GM  \over  r +r_A}  
-{2GM  \over  r}  -{2GMr  \over  r_A^2 }  
\right\} { r_A \over r } L,
\end{equation}
\begin{equation}
  \Delta S \simeq - { \Delta m^2 \over 2E } 
 \left\{
1 - {GM  \over  r +r_A}  -{2GM  \over  r} 
\right\} L,
\end{equation}
where $ L = \sqrt{r^2 - r_A^2 }$ in this case.
The differences increase with the propagating distance $L$,
as expected.
Gravitational corrections are shown in the braces $\{ \}$.  
The coherent wave packet with the initial size $ \sigma $
will survive until $ L \sim E^2 \sigma / \Delta m^2 $.
Gravitationally induced oscillation phase
appears in order $ G \Delta m^2 / E $
except the purely radial motion,
in which the gravitational correction is accidentally canceled 
as seen in eq.(\ref{exs}).
The magnitude of the oscillation phase can be written as 
$\varphi _G \sim \langle \phi\rangle \varphi _0$,
from the dimensional argument.
However, the factor $\langle \phi\rangle $
is not the average of the gravitational potential
over the semi-classical path.
Rather, the correction can be regarded as the modification of the
propagating distance.

\section*{  ACKNOWLEDGMENT }
I would like to thank T. Morozumi for the stimulating conversation.
This work was supported in part
by the Grant-in-Aid for Scientific Research Fund of
the Ministry of Education, Science and Culture of Japan 
(No.08640378).


\end{document}